\documentclass[11pt,twoside]{article}
\pdfoutput=1

\usepackage{wallpaper}

\usepackage{epsfig}
\usepackage{rotate}
\usepackage[figuresright]{rotating}
\usepackage{lscape}
\usepackage{hyperref}
\usepackage{url}
\usepackage{amssymb}
\usepackage{graphicx,caption}
\usepackage{cite}
\usepackage{fancyhdr}
\usepackage{datetime}

\usepackage{multirow}
\usepackage{capt-of}
\usepackage{wrapfig}            
\usepackage{wasysym}            
\usepackage{textcomp}
\DeclareFontFamily{OML}{eur}{\skewchar\font127} \DeclareFontShape{OML}{eur}{m}{n}{<5> <6> 
  <7> <8> <9> gen * eurm <10><10.95><12><14.4><17.28><20.74><24.88>eurm10}{} 
\DeclareSymbolFont{greek}{OML}{eur}{m}{n} 
\DeclareMathSymbol{\upmu}{\mathord}{greek}{"16}

\begin{document}
\title{ILD MC production for detector optimization}

\author{Akiya Miyamoto\footnote{
Talk presented at the International Workshop on Future Linear Colliders (LCWS2018), Arlington, Texas, 22-26 October 2018. C18-10-22.}
\footnote{e-mail: akiya.miyamoto@kek.jp},\\
\small{\it{High Energy Accelerator Research Organization (KEK),}}\\
\small{\it{1-1 Oho, Tsukuba, Ibaraki, 305-0801 Japan,}}\\
\and
~~~~~~~~~~~~~~~~~~~
and
~~~~~~~~~~~~~~~~~~~
\and
Hiroaki Ono\\
\small{\it{Nippon Dental University School of Life Dentistry at Niigata,}} \\
\small{\it{1-8 Hamaura-cho, Chuo-ku, Niigata 951-8580, Japan}}
}

\date{}

\maketitle

\begin{abstract}
A large scale Monte Carlo production has been pursued since spring 2018
for the ILD detector optimization studies based on physics benchmark processes.
A production system based on ILCDirac has been developed to 
produce samples in timely manner. The system and its performance
are presented in this talk.
\end{abstract}


\section{Introduction}
The basic concept of the ILD detector for International Linear Collider was established 
in 2009\cite{ILDLOI} and it has been used extensively to establish physics case of ILC DBD\cite{DBD}
and thereafter. However, the optimumness of the design was concerned in view of progresses in detector technologies and evolution of ILC accelerator design in recent years. To address the question, 
the detector optimization study based on physics benchmark processes based on 
a large scale Monte Carlo production has being conducted.  

Two detector geometries, a large one called ILD\_l5\_v02 
and a small one ILD\_s5\_v02, were selected for comparison and they were
implemented in the DD4hep\cite{DD4hep}  based detector simulation.
In the calorimeter simulation, 
two types of read out for electro-magnetic 
calorimeter, silicon and scintillator, and two types for  
hadronic calorimeter read out, analogue and semi-digital,
were implemented in order to save CPU time and 
clear comparison of model difference.
The simulated data were reconstructed by Marlin data reconstruction framework.

The initial target of the MC production was to produce 500 GeV standard 
model samples which were used in the DBD study\cite{DBD} with 
the same statistics. Differently from the DBD time, the interaction 
vertex points were shifted and smeared depending on 
the initial beam types and low energy $e^{+}/e^{-}$ pairs produced by 
beamstrahlung were overlaid to signal events together with   
lowpt hadron events of large cross section by $\gamma\gamma$ interaction.  
In addition, samples such as single particle events
and quark pair events for detector calibration and basics performance studies
were produced.

Before launching the production, required resources were estimated by 
processing small number of events for all processes. It was found that 
about 0.7M HEPSpec2006 CPU days and about 400 TB of data storage 
would be required to process samples planned. This was not a huge requirements 
for recent GRID infrastructure, however, it was also noticed that 
the reduction of baby sitting works associated with the production 
of more than 600 physics processes required in order to complete 
in timely manner. 
 
\section{Overview of the ILD production system with ILCDirac}

ILD is using DIRAC\cite{DIRAC} and ILCDirac\cite{ILCDIRAC} 
for GRID based MC production after DBD era.
DIRAC provides a high level interface between users and 
GRID resources, such as job managements, file catalogue, transformation 
system for productions, web based interface for monitoring and control.
ILCDirac is an extension for the ILC VO, being developed and operated by 
CLICdp.  The common ILC software and production tools are implemented 
in ILCDirac.  ILD began to use ILCDirac after DBD and ILD MC data produced 
during DBD and thereafter 
have been imported to ILCDirac data catalogue.

In DIRAC system, production is conducted by the transformation system.
When an operator request is received, it selects input files through
database query, creates job scripts, and submit jobs to ILC VO 
cites. It retries job submissions, in case of job failures, several times  
because most of the 
job failure happens due to problematic sites or temporary network 
problem.  Servers for the transformation and workflow are shared among 
ILCDirac users.  The output file name is generated by the transformation workflow module,
which was updated to support the ILD file name convention.

In ILD, the file name was constructed 
by a comma separated list of meta information consists of
one character of key followed by meta value.  The meta information includes
the nominal beam energy and machine parameter, process name, detector model, 
etc. See Ref.~\cite{ILDNamingConvention} for details.
For production purpose, 
file splitting number of generator files and job numbers 
were added in the file name for easy tracking of file creation order.
As a consequence, file name length got close to 128 characters limit of 
ILCDirac and meta values had to be decided carefully 
some times.

In the production, input DBD generator files had to be split (GenSplit)
to smaller number of events
so as to run one job within allowed 
CPU time limit and produced data size within a limit. This task was done 
by a local script. The split files were created in slcio format 
to add header information such as the cross section of the processes
and file sequence number, etc.  The DBD generator files 
were produced with stdhep format which did not allow to store 
additional header information. In splitting, the number of events for each file 
were determined so that the reconstruction data size 
per job be within O(1GB) and CPU time for simulation be less than 
O(8 hour) depending on input generator processes. 

The DBD samples at 500 GeV CM energy includes events not only by 
$e^+e^-$ collision but also $\gamma\gamma$ and $e\gamma$; 
final states from 2-fermion to 6-fermion.  There were more than 600 processes,
some of them could complete
within a little CPU time; some consumed quite a lot of CPU time. 
In order to produce them efficiently, processes were grouped to about 
50 groups considering the initial beam nature
( IP smearing parameters and type of background to be overlaid ).
Number of jobs per transformation were also limited to be O(10k).
The baby sitting work of production were made based on group.

After the one group of production, DST merge jobs were executed 
using ILCDirac UserJob tool.  The DST files, which have typically several 
to a few tens of MB, were collected based on the physics process name 
and ID and merged keeping the order of file sequence in GenSplit.

\begin{figure}[htbp]
\begin{center}
  \includegraphics[width=0.7\columnwidth]{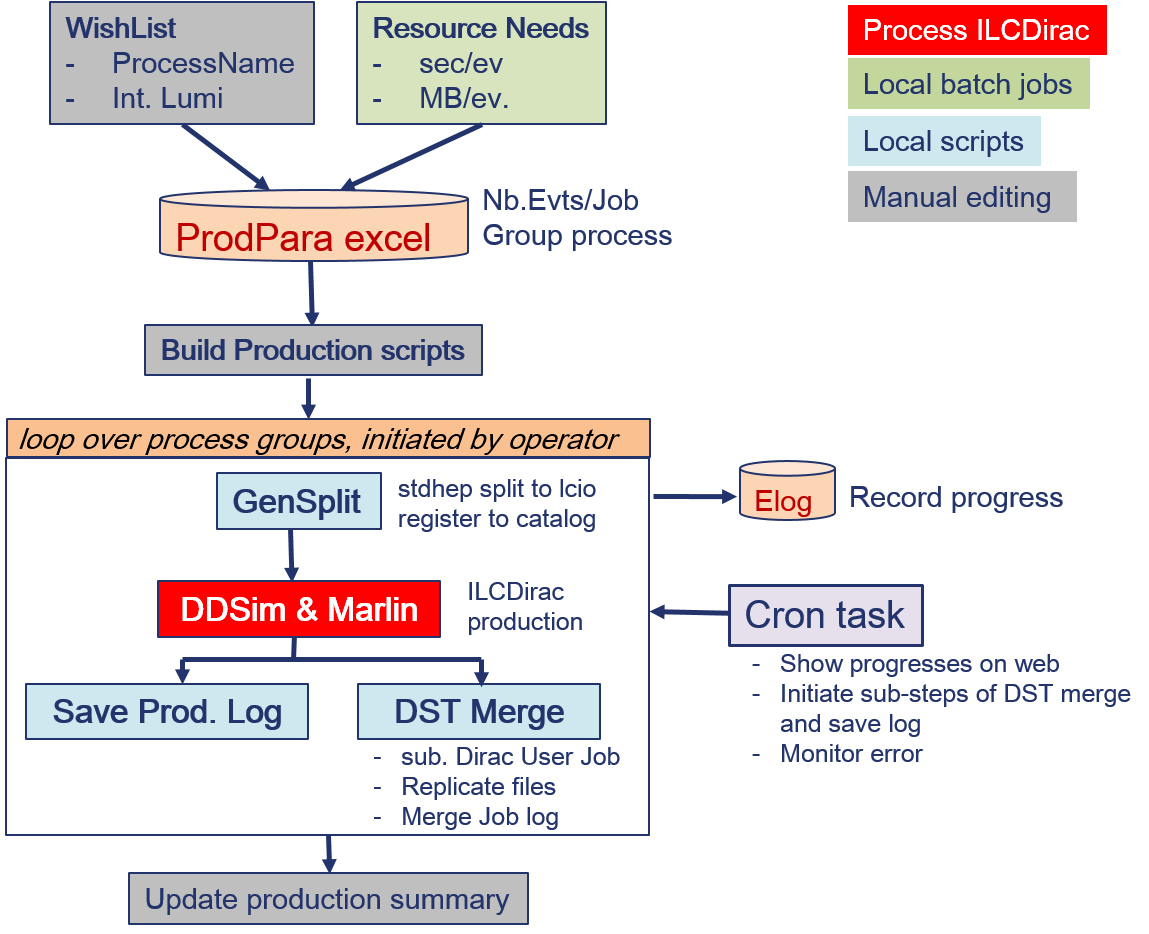}
\end{center}
 \captionsetup{width=0.8\textwidth}
 \caption{\it The workflow of ILD MC Production. The task done by the DIRAC transformation is shown in red box, scripts at local host in cyan box, 
local batch jobs in green box, interactive tasks in grey box.
 \label{Fig:ILDMCProductionWorkflow}}
\end{figure}
Overall workflow of the ILD MC production is shown in Figure~\ref{Fig:ILDMCProductionWorkflow}.
In the Figure, the task done by ILCDirac production is shown by a red box,
tasks done by ILCDirac user job tools are indicated by cyan boxes. 
Prior to start production, a ProdPara excel file (shown by pink disk) was created 
based on the production WishList and the Resource Needs estimation. 
It was used to group the generator processes and build a chain of scripts to perform GenSplit, DDSim and Marlin production, DSTMerge and Saving log for each group.
The chain of scripts were monitored by a cron script, which also initiated jobs for 
DST merge and save log automatically.

ELOG system\cite{ELOGReference} was used to keep a track of DIRAC transformation and 
a database for the produced samples. ELOG uses ASCII data records, thus
easy to use and flexible. A simple search function provided is also useful. Python interface 
to ELOG were utilized for automatic logging of production progresses and status.   
Our ELOG server is available at \url{https://ild.ngt.ndu.ac.jp/elog}. This server is also used 
for the database of generator samples produced since DBD era. Files  
of all generator meta data in json format is also provided.

\section{Production results}
\begin{figure}[htbp]
\begin{center}
  \includegraphics[width=0.7\textwidth]{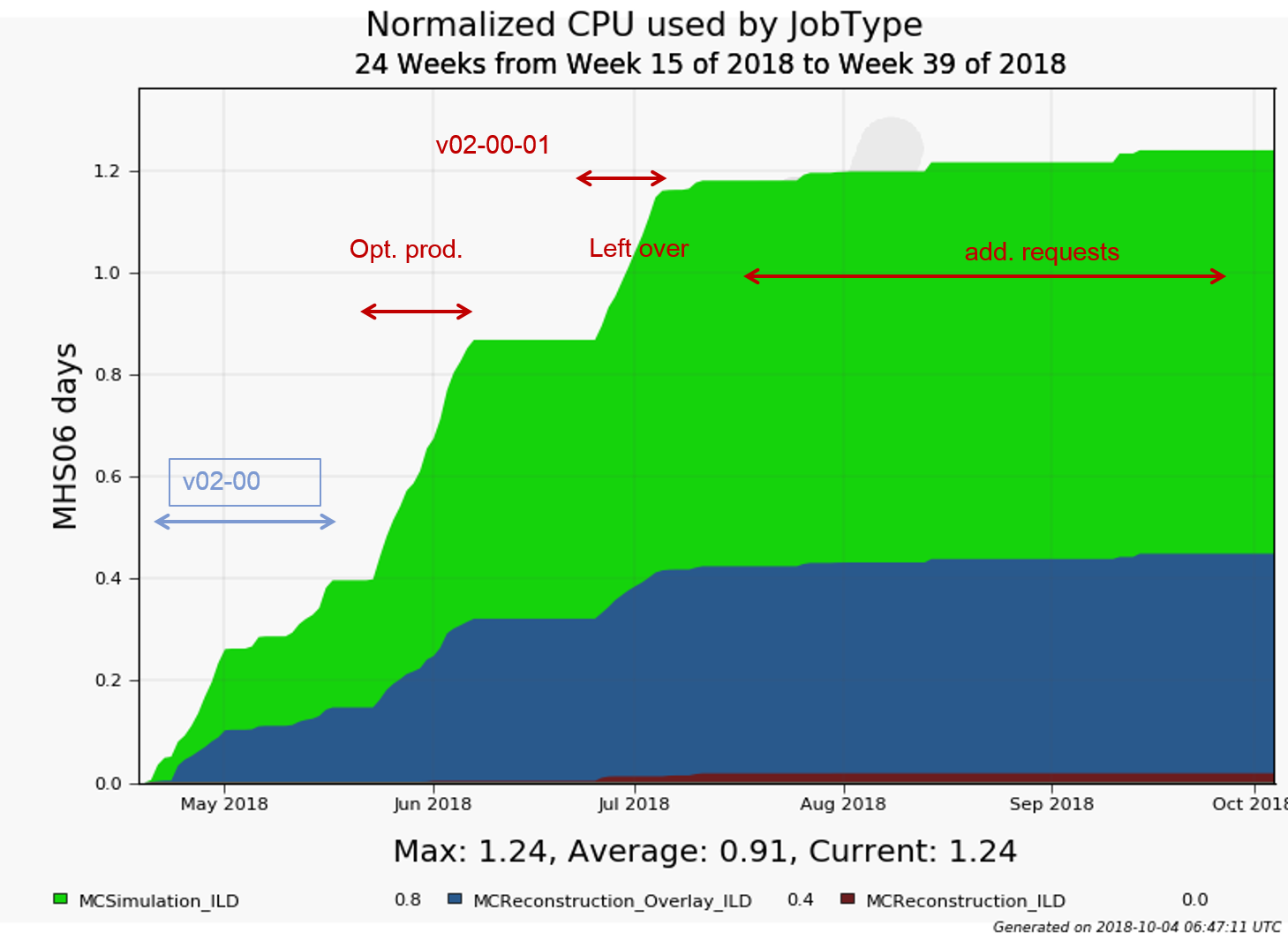}
\end{center}
 \captionsetup{width=0.8\textwidth}
  \caption{\it Cumulative normalized CPU in unit of  Mega HEPSpecInt2006 units,
  since begining of the opimization production.
  \label{Fig:CumulativeNormalizedCPU}}
\end{figure}

The optimization MC production was launched late April 2018, 
using ILCSoft version v02-00. Unfortunately, the problem in v02-00 
was found and new production was launched again late May 2018
with v02-00-01 ILCSoft.  Since then, the production proceeded smoothly 
and most of the planned events could be produced within a month.
Production of left over samples and additional samples continued 
time to time since then. This could be seen in Figure~\ref{Fig:CumulativeNormalizedCPU},
which shows the cumulative normalized CPU used since 
April 2018.

\begin{figure}[htbp]
\begin{center}
  \includegraphics[width=0.48\textwidth]{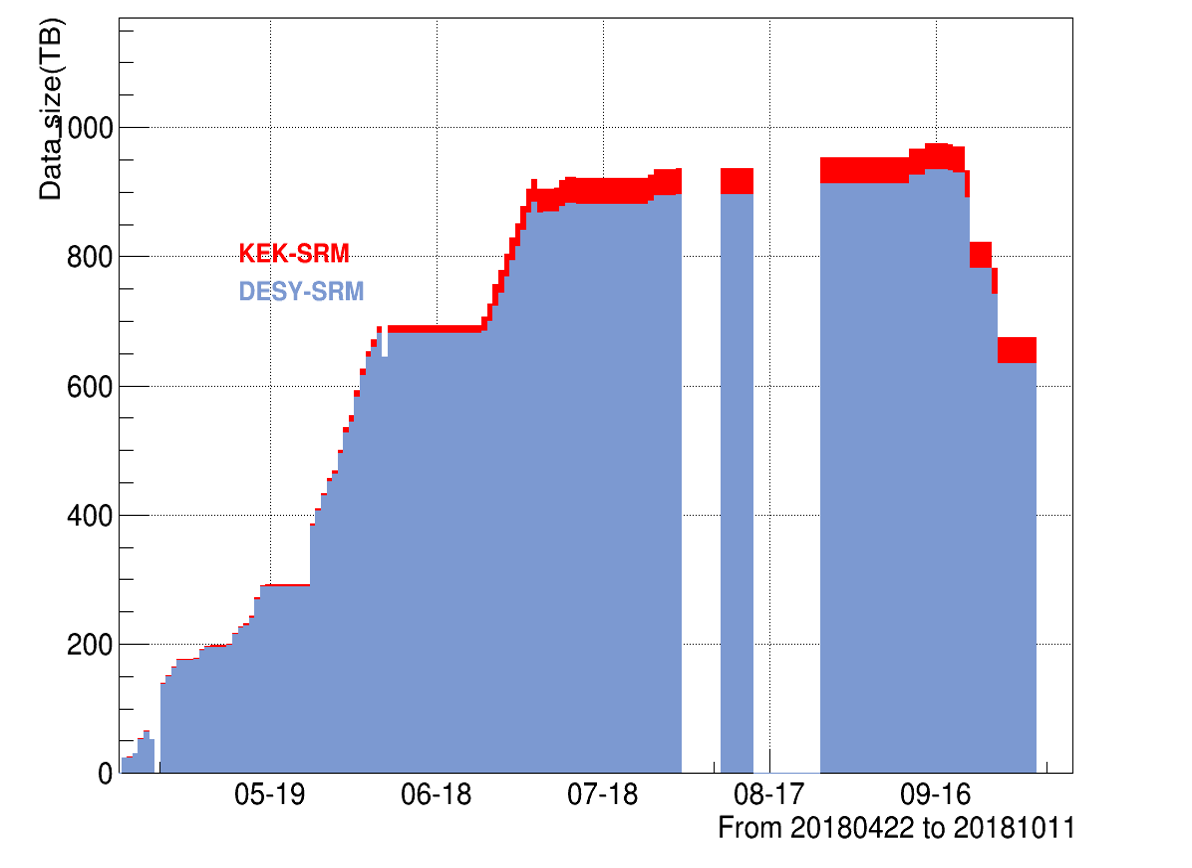}
  \hspace{0.02\textwidth}
  \includegraphics[width=0.48\textwidth]{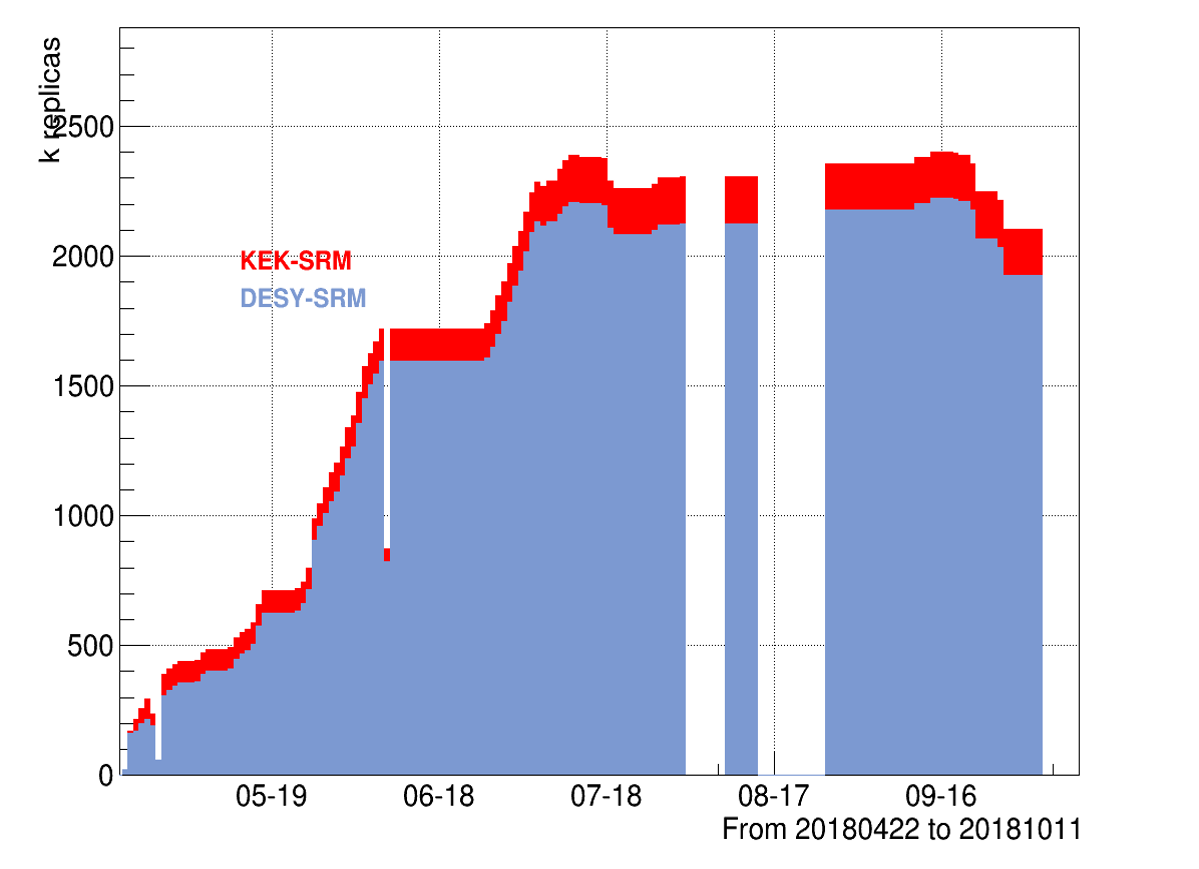}
\end{center}
 \captionsetup{width=0.8\textwidth}
  \caption{\it Produced data size (left) and number of replicas (right) since middle of April 2018
  \label{Fig:DataSizeAndReplicas}}
\end{figure}
The size and the number of data produced this period is shown in 
Figure~\ref{Fig:DataSizeAndReplicas}.  In our production, all produced data, namely 
simulated, reconstructed, DST, and DSTMerged data were kept and stored 
at DESY. DST-merged files and small part of files produced for calibration 
were replicated to KEK.  Erroneous data produced by  ILCSoft v02-00 were 
removed in September to save about 300TB of storage space.

\begin{figure}[htbp]
\begin{center}
  \includegraphics[width=0.48\textwidth]{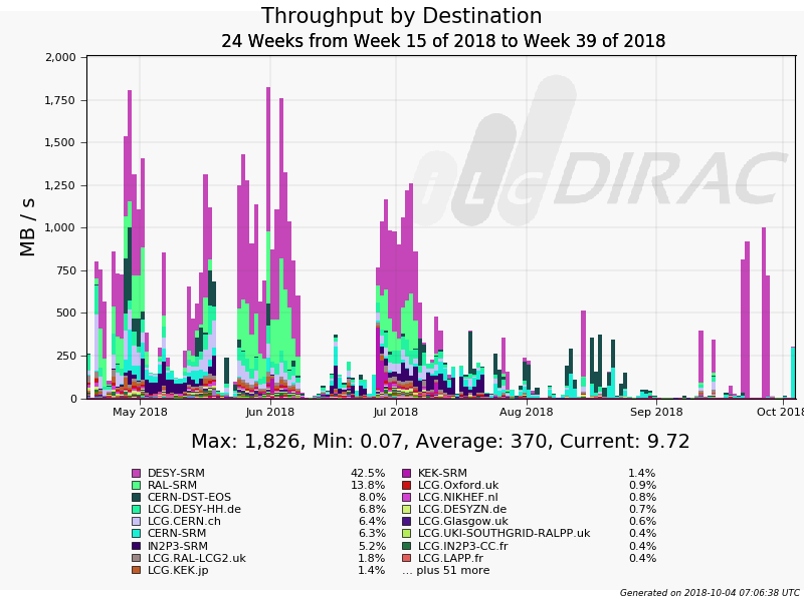}
  \hspace{0.02\textwidth}
  \includegraphics[width=0.48\textwidth]{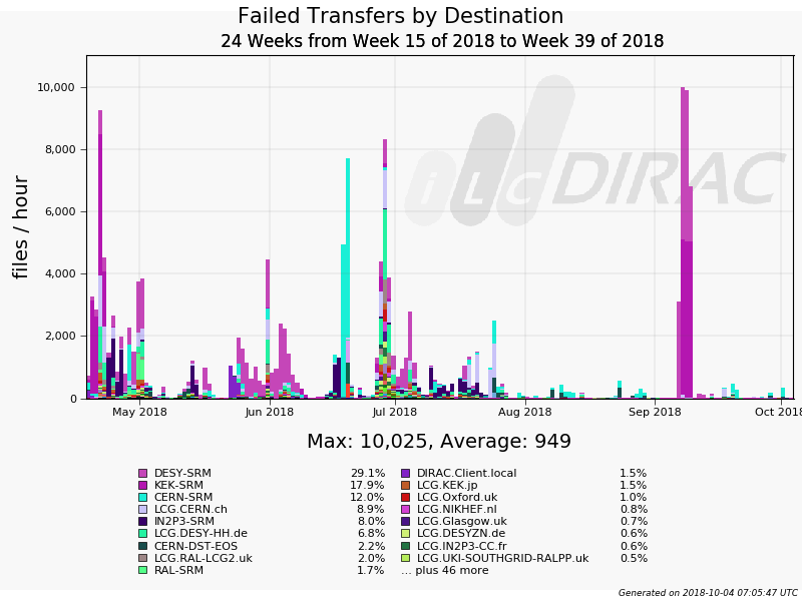}
\end{center}
 \captionsetup{width=0.8\textwidth}
  \caption{
  \it Day by day records of throughput by destination (left) and failed transfer by destination (right).
  \label{Fig:TransferAndFailedRate}}
\end{figure}

Since DESY was the primary storage, it's performance affected error rates of production.
The rate of the all transfer and failed transfer to DESY monitored by ILCDirac 
are shown in Figure~\ref{Fig:TransferAndFailedRate}. It is seen that 
the data transfer rate exceeded 1.5GB/sec at peak and 
the error rates correlated to the transfer rate. 
In order to avoid high error rate, the number of jobs in the ILCDirac was 
restricted to less than about 5000 by 
limiting the number of gensplit jobs running in parallel or 
the number of tasks assigned to the transformation. Former method limited 
the number of new input files for the simulation jobs, thus restricted the job 
submission by transformation.

\section{Summary}
A tool for ILD MC production has been developped using ILCDirac.
It has been used for the production of MC samples for ILD detector optimization. 
About ~1 PB of data has been produced since May 2018 and met the needs for the ILD detector optimization studies.
Produced samples were mostly 500 GeV samples with ILD large and small geometries.
Relatively small amount of samples of other energies, 91 GeV and 1000 GeV, and those for detector calibration and basic performance studies were produced as well. 
The ILD production tools\cite{ILDProdRepo} worked smoothly and  
baby sitting works, which had been the major limitation in previous test productions ( summer 2017-winter 2018 ), were reduced significantly. 
Currently, the data transfer rate is the major limiting factor of the production rate.
The refinement of the production scripts are in progress, aiming for the use in coming 250 GeV production.

\section*{Acknowledgements}
The authors would like to thank Andre Sailer and Marco Petri\v{c} for supporting ILCDirac for the MC production, Frank Gaede and ILD Software group members for discussion and comments during the course of this work.
\bibliographystyle{utphys}
\bibliography{ILDMCProduction}

\end{document}